\documentclass[pra,showpacs,superscriptaddress,hyperref,twocolumn,bibnotes]{revtex4-1}
\usepackage{amsfonts}
\usepackage{hyperref}
\usepackage{amsmath}
\usepackage{amssymb}
\usepackage{graphicx}

\providecommand{\U}[1]{\protect\rule{.1in}{.1in}}

\begin{document}

\title{Equivalence between Redfield and master equation approaches for a
time-dependent quantum system and coherence control}
\author{D. O. Soares-Pinto}
\affiliation{Instituto de F\'{\i}sica de S\~{a}o Carlos, Universidade de
S\~{a}o Paulo, Caixa Postal 369, 13560-970 S\~{a}o Carlos, S\~{a}o Paulo,
Brazil}
\author{M. H. Y. Moussa}
\affiliation{Instituto de F\'{\i}sica de S\~{a}o Carlos, Universidade de
S\~{a}o Paulo, Caixa Postal 369, 13560-970 S\~{a}o Carlos, S\~{a}o Paulo,
Brazil}
\author{J. Maziero}
\affiliation{Centro de Ci\^{e}ncias Naturais e Humanas, Universidade Federal
do ABC, R. Santa Ad\'{e}lia 166, 09210-170 Santo Andr\'{e}, S\~{a}o Paulo,
Brazil}
\author{E. R. deAzevedo}
\affiliation{Instituto de F\'{\i}sica de S\~{a}o Carlos, Universidade de
S\~{a}o Paulo, Caixa Postal 369, 13560-970 S\~{a}o Carlos, S\~{a}o Paulo,
Brazil}
\author{T. J. Bonagamba}
\affiliation{Instituto de F\'{\i}sica de S\~{a}o Carlos, Universidade de
S\~{a}o Paulo, Caixa Postal 369, 13560-970 S\~{a}o Carlos, S\~{a}o Paulo,
Brazil}
\author{R. M. Serra}
\affiliation{Centro de Ci\^{e}ncias Naturais e Humanas, Universidade Federal
do ABC, R. Santa Ad\'{e}lia 166, 09210-170 Santo Andr\'{e}, S\~{a}o Paulo,
Brazil}
\author{L. C. C\'{e}leri}
\affiliation{Centro de Ci\^{e}ncias Naturais e Humanas, Universidade Federal
do ABC, R. Santa Ad\'{e}lia 166, 09210-170 Santo Andr\'{e}, S\~{a}o Paulo,
Brazil}
\email{lucas.celeri@ufabc.edu.br}


\begin{abstract}
We present a derivation of the Redfield formalism for treating the
dissipative dynamics of a time-dependent quantum system coupled to a
classical environment. We compare such a formalism with the master equation
approach where the environments are treated quantum mechanically. Focusing
on a time-dependent spin-$1/2$ system we demonstrate the equivalence between
both approaches by showing that they lead to the same Bloch equations and,
as a consequence, to the same characteristic times $T_{1}$ and $T_{2}$
(associated with the longitudinal and transverse relaxations, respectively).
These characteristic times are shown to be related to the operator-sum
representation and the equivalent phenomenological-operator approach.
Finally, we present a protocol to circumvent the decoherence processes due
to the loss of energy (and thus, associated with $T_{1}$). To this end, we
simply associate the time-dependence of the quantum system to an easily
achieved modulated frequency. A possible implementation of the protocol is
also proposed in the context of nuclear magnetic resonance.
\end{abstract}

\maketitle

\section{Introduction}

The rapid development of quantum information science has brought together
several areas of theoretical and experimental physics \cite{Plataforms}.
Much effort has been concentrated in the search for solutions to sensitive
problems that prevent the efficient realization of quantum information
processing \cite{Nielsen}. We first mention the system-environment coupling
which induces the decoherence of quantum states \cite{Decoherence}, apart
from other barriers such as scalability \cite{Scalability} and optimal
control of individual systems \cite{Optimal}. These challenges motivate both
fundamental physical phenomena and outstanding technological issues such as
individually addressing quantum systems, separated by only few $n$m, with
small errors \cite{Nagerl}.

Potential platforms for the implementation of quantum logic operations
appeared in many fields such as condensed matter, quantum optics, and atomic
physics \cite{Plataforms}. However, the problems mentioned above are faced
by all the different communities when employing their particular techniques.
In the particular case of the dissipation and decoherence phenomena ---in
which we focus in the present work--- the Redfield formalism \cite%
{Redfield,Slichter} and the master equation \cite{SZ} have been the most
applied approaches to address the environment effects on the proposed
protocols for quantum information processing. Whereas the semiclassical
Redfield formalism relies on a classical noise source, a quantum environment
is assumed in the master equation approach. In this article, considering the
general case of a time-dependent system, we discuss general similarities and
differences between both approaches and show that they are equivalent, in
the sense that they lead to the same phenomenological Bloch equations \cite%
{Bloch}. Consequently, both of them result in the same characteristic
relaxation times $T_{1}$ and $T_{2}$ associated with the longitudinal and
transverse relaxations, respectively \cite{Redfield, Slichter}. From this
identification we show how these characteristic times are related to the
operator-sum representation \cite{Nielsen} and the phenomenological-operator
approach \cite{Miled}.

The Redfield formalism was intended to offer a microscopic description of
the relaxation phenomenon, thus providing a deeper understanding of the
parameters $T_{1}$ and $T_{2}$. Whereas the classical noise source employed
in the Redfield theory suffices to derive both relaxation times, two
distinct quantum environments must be adopted to derive these time scales
from the master equation formalism. On this regard, an amplitude and a phase
damping environment are assumed to define the longitudinal and the
transverse relaxation times, respectively. These quantum environments
represent an energy-draining and a phase-shuffle channel by which the system
loses excitations and phase relations.

After presenting a detailed derivation of the Redfield theory and comparing
the derived characteristics times with those obtained from the master
equation, we finally apply these equivalent formulations to the problem of
state protection. We note that several distinct techniques have been
proposed to control the effects of decoherence on quantum states, aiming to
enlarge the fidelity of quantum information protocols. Among others, we
mention the quantum-error correction codes \cite{QECC}, environments
engineering \cite{ER}, decoherence-free subspaces \cite{DFS}, and dynamical
decoupling \cite{DD}. We finally mention that in a previous work \cite{Lucas}%
, addressing the energy draining and decoherence of a harmonic oscillator,
it was demonstrated that the inevitable action of the environment can be
substantially weakened when considering appropriate non-stationary quantum
systems. Reasoning by analogy with the technique presented in Ref. \cite%
{Lucas}, we show how to enlarge the longitudinal relaxation time associated
with the amplitude-damping channel focusing a spin-$1/2$ system. The ideas
presented here for decoherence control can be easily implemented in the
nuclear magnetic resonance (NMR) context.

This article is organized as follows: In Sec. II we present the derivation
of the Redfield equation for the general scenario of a time-dependent
system. In Sec. III we apply the master equation approach to the same case.
In Sec. IV we show the equivalence between the Redfield and the master
equation formalisms by deriving the Bloch equations from both approaches.
Focusing on a time-dependent spin-$1/2$ system, in Secs. V and VI we present
the operator-sum representation and the phenomenological operator approaches
and their relation with the previous techniques. As an application of the
theory, in Sec. VII we address the state protection of a non-stationary spin-%
$1/2$ system. Finally, Sec. VIII is dedicated to our final remarks where we
discuss the generalization of the methods presented in this article for
larger systems. In general, we will adopt the language of the NMR quantum
information processing \cite{NMR}, although the theory presented here is
valid for several other platforms, as quantum dots \cite{Qdots},
superconducting artificial atoms \cite{SAA}, etc. Throughout the article we
will use natural units such that $\hbar =k_{B}=1$.

\section{Redfield formalism for a time-dependent spin system}

Considering the interaction of a time-dependent spin system, described by
the Hamiltonian $H_{S}(t)$, with a spin lattice modelling the environment
and represented by the Hamiltonian $H_{L}$, the total density operator $\rho
^{sch}(t)$ in the Schr\"{o}dinger picture, evolves as 
\begin{equation}
\frac{d\rho ^{sch}(t)}{dt}=-i\left[ H_{S}(t)+H_{L}+H_{SL}(t),\rho ^{sch}(t)%
\right] ,  \label{1}
\end{equation}%
$H_{SL}(t)$ being the time-dependent spin-lattice interaction. In the
interaction picture, defined by the unitary transformation $U(t)=\exp
\left\{ -i\left( \mathcal{H}_{S}(t)+H_{L}t\right) \right\} $, where $%
\mathcal{H}_{S}(t)=\int_{0}^{t}dt^{\prime }H_{S}(t^{\prime })$, we simplify
the above evolution equation to the form 
\begin{equation}
\frac{d\rho (t)}{dt}=-i\left[ V_{SL}(t),\rho (t)\right] ,  \label{2}
\end{equation}%
where $V_{SL}(t)=e^{i\mathcal{H}_{S}(t)}\mathcal{H}_{SL}(t)e^{-i\mathcal{H}%
_{S}(t)}$, with $\mathcal{H}_{SL}(t)=e^{iH_{L}t}H_{SL}(t)e^{-iH_{L}t}$. We
have assumed the condition $\left[ H_{S}(t),H_{S}(t^{\prime })\right] =0$
which is always fulfilled whenever the free Hamiltonian of the system can be
written as a diagonal time-independent operator, with a time-dependent
coefficient $H_{S}\left( t\right) =f\left( t\right) O_{S}$\textbf{.}
Moreover, we observe that in NMR relaxation experiments all the required
pulses to perform the necessary rotations are applied either at the
beginning, to prepare the initial state $\rho ^{sch}(0)$, or at the end of
the experiment, to implement the tomography of the evolved state. Between
the applications of these pulses, the prepared state $\rho ^{sch}(0)$ of the
system, described by the diagonal Hamiltonian $H_{S}(t)$, evolves only under
the action of the environment. 

By its turn, the density operator in the rotating frame is given by $\rho
(t)=e^{i\mathcal{H}_{S}(t)}\widetilde{\rho }^{sch}(t)e^{-i\mathcal{H}_{S}(t)}
$ where $\widetilde{\rho }^{sch}(t)=e^{iH_{L}t}\rho ^{sch}(t)e^{-iH_{L}t}$.
By assuming a weak system-environment coupling and getting rid of the
degrees of freedom of the spin lattice, we solve Eq. (\ref{2}) up to second
order of perturbation theory, obtaining 
\begin{equation*}
\frac{d\sigma (t)}{dt}=-\mbox{Tr}_{L}\int_{0}^{t}dt^{\prime }\left[
V_{SL}(t),\left[ V_{SL}(t^{\prime }),\rho (t^{\prime })\right] \right] .
\end{equation*}%
We considered that the interaction $V_{SL}(t)$ is a stochastic operator with
null mean value \cite{Redfield, Slichter}, which results in a zero first
order term. Next, let us use the Markov approximation $\rho (t^{\prime
})\rightarrow \rho (t)\simeq \sigma (t)\otimes \rho _{L}(0)$, $\sigma (t)$
being the reduced density operator of the spin system, $\sigma (t)=\mbox{Tr}%
_{L}\rho (t)$, and $\rho _{L}$ the reduced density operator of the
environmental spin lattice. This approximation means that the state of the
lattice is not affected by the interaction with the system. In other words,
it means that the lattice presents a sufficiently large heat capacity in
order to remain in the thermal equilibrium state $\rho _{L}(0)=e^{-\beta
H_{L}}/\mbox{Tr}\left[ e^{-\beta H_{L}}\right] $, with $\beta =1/T$, $T$
being the environment temperature. Finally, inspired in NMR systems \cite%
{Slichter, NMR}, we are going to apply the high-temperature approximation,
which takes into consideration systems were the energy gap between the spin
levels, $\hbar \omega $ (where $\omega $ is the characteristic transition
frequency among levels), is much smaller than the thermal energy, $k_{B}T$,
of the system, i.e. $\hbar \omega /k_{B}T\ll 1$. In this sense, the density
operator of the system can be written as $\sigma (t)=e^{-\beta
H_{S}(t)}/Z\simeq 1-\beta H_{S}(t)$. We stress that there is a crucial
difference between high and infinite temperature limits; differently from
the latter case, in the former there is still a population difference
between the spin levels which accounts for the reminiscent  equilibrium
magnetization. Thus, applying the high-temperature approximation we obtain $%
\sigma (t)\otimes \rho _{L}(0)\simeq \sigma (t)-\beta H_{L}$ and,
consequently 
\begin{equation}
\frac{d\sigma (t)}{dt}=\mbox{Tr}_{L}\int_{0}^{t}dt^{\prime }\left[ V_{SL}(t),%
\left[ V_{SL}(t^{\prime }),\beta H_{L}-\sigma (t)\right] \right] .  \label{3}
\end{equation}%
%
From the Heisenberg equation of motion for $V_{SL}(t)$, we obtain 
\begin{align*}
\left[ V_{SL}(t),\sigma (t)-\beta H_{L}\right] =& \left[ V_{SL}(t),\sigma
(t)+\beta H_{S}(t)\right]  \\
& -i\beta \frac{dV_{SL}(t)}{dt} \\
& +i\beta U^{\dagger }(t)\frac{dH_{SL}(t)}{dt}U(t),
\end{align*}%
and, consequently, the evolution equation reads 
\begin{align}
& \frac{d\sigma (t)}{dt}=i\beta \mbox{Tr}_{L}\left\{ \left[
V_{SL}(t),V_{SL}(0)\right] \right\}   \notag \\
& -\mbox{Tr}_{L}\int_{0}^{t}dt^{\prime }\left[ V_{SL}(t),\left[
V_{SL}(t^{\prime }),\sigma (t)+\beta H_{S}(t^{\prime })\right] \right]  
\notag \\
& -i\beta \mbox{Tr}_{L}\int_{0}^{t}dt^{\prime }\left[ V_{SL}(t),U^{\dagger
}(t^{\prime })\frac{dH_{SL}(t^{\prime })}{dt^{\prime }}U(t^{\prime })\right]
.  \label{4}
\end{align}%
By rewriting the spin-lattice interaction as $V_{SL}(t)\propto \lambda (t)%
\mathcal{O}$, where $\lambda (t)$ models the lattice stochastic fluctuation
and $\mathcal{O}$ stems for an operator acting on the spin system space, we
verify straightforwardly that the first term of the r.h.s. of Eq. (\ref{4})
is null, in accordance with the assumption $\left\langle \lambda
(t)\right\rangle =0$. Moreover, with the above definition for the
spin-lattice interaction, we verify that $\mathcal{H}_{SL}(t)=H_{SL}(t)$
and, consequently, $V_{SL}(t)=e^{i\mathcal{H}_{S}(t)}H_{SL}(t)e^{-i\mathcal{H%
}_{S}(t)}$. Integrating by parts the third term of the r.h.s. of Eq. (\ref{4}%
) and considering, as usual, that the time oscillations of the operator $%
U^{\dagger }(t^{\prime })\mathcal{O}U(t^{\prime })$ is much faster than that
of $V_{SL}(t)$, we apply the rotating wave approximation to conclude that
this term is also null. The fact that this is indeed the case, can be seen
as follows: the operator $V_{SL}(t)$ oscillates with the spin-lattice
coupling frequency, while the operator $U^{\dagger }(t^{\prime })\mathcal{O}%
U(t^{\prime })$ oscillates with the bare spin frequency which (in the
assumed system-environment weak coupling regime) is much higher than the
interaction frequency. Putting all this together, we finally obtain the
simplified equation of motion for the spin system 
\begin{equation*}
\frac{d\sigma (t)}{dt}=\mbox{Tr}_{L}\int_{0}^{t}dt^{\prime }\left[ V_{SL}(t),%
\left[ V_{SL}(t^{\prime }),\beta H_{S}(t^{\prime })-\sigma (t)\right] \right]
,
\end{equation*}%
which, in accordance with the high-temperature approximation, where $\sigma
_{eq}(0)\simeq 1-\beta H_{S}(0)$ and $\beta H_{S}(t^{\prime })\simeq
1-\sigma _{eq}(0)$, becomes 
\begin{equation}
\frac{d\Sigma (t)}{dt}=-\int_{0}^{t}dt^{\prime }\overline{\left[ V_{SL}(t),%
\left[ V_{SL}(t^{\prime }),\Sigma (t)\right] \right] },  \label{5}
\end{equation}%
where we have defined the operator $\Sigma (t)=\sigma (t)-\sigma _{eq}(0)$
and substituted the trace over the lattice degrees of freedom by the
ensemble average over stochastic realizations, represented by the over bar.
We have thus obtained the Redfield equation for a time-dependent spin system
and we note, in spite of the $c$-number character of the environment degrees
of freedom, its resemblance with the master equation to be presented below. 

It is important to stress that the high-temperature approximation, allowing
us to define the latter operator $\Sigma (t)$, indicates a relaxation to the
highly mixed thermal Gibbs state. However, we mention that in the whole
calculation to obtain the Redfield equation (\ref{5}) it is not necessary to
impose such approximation. It was only done because it is characterisitic of
NMR systems, on which we focus in the present work.

Towards the definition of the lattice spectral density we next introduce,
through the eigenvalue equation $H_{S}(t)\left\vert k\right\rangle
=\epsilon_{k}(t)\left\vert k\right\rangle $, the spin basis $\left\{
\left\vert k\right\rangle \right\} $. Taking the matrix element $kk^{\prime
} $ of Eq. (\ref{5}) and back to Schr\"{o}dinger picture where $\Sigma
^{sch}(t)=e^{-i\mathcal{H}_{S}(t)}\Sigma (t)e^{i\mathcal{H}_{S}(t)}$, we
obtain the Redfield equations for the evolution of the density matrix
elements 
\begin{align}
\frac{d\Sigma _{kk^{\prime }}^{sch}(t)}{dt}& =-i\left\langle k\right\vert %
\left[ H_{S},\Sigma ^{sch}\right] \left\vert k^{\prime }\right\rangle  \notag
\\
& +\sum_{n,n^{\prime }}e^{-i\left( \Omega _{kk^{\prime }}+\Omega_{n^{\prime
}n}\right) }R_{kn,n^{\prime }k^{\prime }}\Sigma _{nn^{\prime}}^{sch},
\label{6}
\end{align}
where we have used the short-hand notation $\Omega_{kn}(t)=E_{k}(t)-E_{n}(t) 
$, $E_{k}(t)=\int_{0}^{t}dt^{\prime }\epsilon_{k}(t^{\prime })$, $\Sigma
_{nn^{\prime }}^{sch}(t)=\left\langle n\right\vert \Sigma
^{sch}(t)\left\vert n^{\prime }\right\rangle $, and defined the relaxation
matrix elements 
\begin{align}
R_{kn,n^{\prime }k^{\prime }}(t)=& J_{kn,n^{\prime }k^{\prime
}}(t,\Omega_{n^{\prime }k^{\prime }})e^{i\Omega _{kn}(t)}  \notag \\
& +J_{n^{\prime }k^{\prime },kn}(t,\Omega _{kn})e^{i\Omega
_{n^{\prime}k^{\prime }}(t)}  \notag \\
& -\delta _{k^{\prime }n^{\prime
}}\sum_{j}J_{kj,jn}(t,\Omega_{jn})e^{i\Omega _{kj}(t)}  \notag \\
& -\delta _{kn}\sum_{j}J_{jk^{\prime },n^{\prime }j}(t,\Omega_{n^{\prime
}j})e^{i\Omega _{jk^{\prime }}(t)},  \label{7}
\end{align}
with the environment spectral densities given by 
\begin{subequations}
\label{8}
\begin{align}
J_{kn,n^{\prime }k^{\prime }}(t,\Omega _{n^{\prime }k^{\prime }}) &
=\int_{0}^{t}dt^{\prime }G_{kn,n^{\prime }k^{\prime
}}(t,t^{\prime})e^{i\Omega _{n^{\prime }k^{\prime }}(t^{\prime })},
\label{8a} \\
G_{kn,n^{\prime }k^{\prime }}(t,t^{\prime })& =\overline{\left\langle
k\right\vert H_{SL}(t)\left\vert n\right\rangle \left\langle
n^{\prime}\right\vert H_{SL}(t^{\prime })\left\vert k^{\prime }\right\rangle 
}.  \label{8b}
\end{align}
To simplify the notation we have omitted the explicit time dependence of all
functions in Eq. (\ref{6}).

For the particular case of a time-independent system, we obtain $%
E_{m}(t)=\epsilon _{m}t$, $\Omega _{kn}(t)=\left( \epsilon
_{k}-\epsilon_{n}\right) t\equiv \omega _{kn}t$ and the above Redfield
equations reduce to the well-know text book result \cite{Redfield, Slichter} 
\end{subequations}
\begin{align}
\frac{d\Sigma _{kk^{\prime }}^{sch}(t)}{dt}=& -i\left\langle k\right\vert %
\left[ H_{S},\Sigma ^{sch}(t)\right] \left\vert k^{\prime }\right\rangle 
\notag \\
& +\sum_{n,n^{\prime }}e^{-i\left( \omega _{kk^{\prime }}-\omega_{n^{\prime
}n}\right) t}R_{kn,n^{\prime }k^{\prime }}(t)\Sigma _{nn^{\prime}}^{sch}(t),
\notag
\end{align}
with 
\begin{align*}
R_{kn,n^{\prime }k^{\prime }}=& J_{kn,n^{\prime }k^{\prime
}}(\omega_{n^{\prime }k^{\prime }})e^{i\omega _{kn}t}+J_{n^{\prime
}k^{\prime},kn}(\omega _{kn})e^{i\omega _{n^{\prime }k^{\prime }}t} \\
& -\delta _{k^{\prime }n^{\prime }}\sum_{j}J_{kj,jn}(\omega_{jn})e^{i\omega
_{kj}t} \\
& -\delta _{kn}\sum_{j}J_{jk^{\prime },n^{\prime }j}(\omega_{nj})e^{i\omega
_{jk^{\prime }}t},
\end{align*}
and 
\begin{equation*}
J_{kn,nn^{\prime }}(\omega _{nn^{\prime }})=\int_{0}^{\infty}dt^{\prime
}G_{kn,nn^{\prime }}(t^{\prime })\exp \left\{ i\omega_{nn^{\prime
}}t^{\prime }\right\} .
\end{equation*}

We observe that, although we have focused on a spin system, the equations
obtained here are completely general, being valid for whichever the
Hamiltonian $H_{S}(t)$, provided that the three following conditions are
met: $i)$ $[H_{S}(t),H_{S}(t^{\prime })]=0$, $ii)$ system-environment weak
coupling regime (Markovian environment), and iii) high-temperature
approximation (specifically, to derive Eqs. (\ref{6}), (\ref{7}), and (\ref%
{8})). The restrictions and the validity of these approximations will be
discussed in the conclusions of the article. Let us now turn to the master
equation approach.

\section{The master equation approach}

In this section, in contrast to the semiclassical approach of the Redfield
formalism, we derive the master equation governing the dynamics of the
dissipative time-dependent spin system where the environment is assumed to
be modelled within the quantum formalism. We start from Eq. (\ref{3}), such
that 
\begin{equation}
\frac{d\sigma (t)}{dt}=-\mbox{Tr}_{L}\int_{0}^{t}dt^{\prime }\left[
V_{SL}(t),\left[ V_{SL}(t^{\prime }),\sigma (t)\right] \right] .  \label{9}
\end{equation}
Now, instead of assuming a classical environment leading to the above
defined spin-lattice interaction as $V_{SL}(t)\propto \lambda (t)\mathcal{O}$%
, we consider two distinct quantum environments, to be defined below as the
amplitude- and the phase-damping channels, each one being modelled by an
infinite collection of decoupled harmonic oscillators, described by the
Hamiltonian $H_{L}=\sum_{r,\ell }$ $\varpi _{r\ell }a_{r\ell
}^{\dagger}a_{r\ell }$ where $r=1,2$ labels the environments while $\ell$
stands for the infinity set of oscillators whose frequencies are denoted by $%
\varpi_{r\ell }$. $a_{r\ell }^{\dagger }$ ($a_{r\ell }$) represents the
creation (annihilation) operator for the $l$th mode of the $r$th
environment. The action of these environments on the spin system is modelled
by the interaction 
\begin{equation}
V_{SL}(t)=\sum\limits_{r}\left[ \mathcal{O}_{r}^{\dagger }\Gamma _{r}(t) + 
\mathcal{O}_{r}\Gamma _{r}^{\dagger }(t)\right] ,  \label{10}
\end{equation}
where $\Gamma _{r}(t)=\sum\nolimits_{\ell }\gamma _{r\ell }(t)a_{r\ell }$
and $\gamma _{r\ell }(t)=$ $\gamma _{r\ell }^{Schr}(t)\exp \left[
i\Delta_{k}(t)\right] $ with $\Delta _{k}(t)$ being the phase factor coming
from the transformation $U^{\dagger }(t)\mathcal{O}_{r}^{\dagger }a_{r\ell
}U(t)$ to the interaction picture. It is worth mentioning that the
time-dependence of the system-environment coupling in the Schr\"{o}dinger
picture, $\gamma_{r\ell }^{Schr}(t)$, comes from the assumption of a
time-dependent spin system Hamiltonian, $H_{S}(t)$. In fact, the coupling
strength $\gamma_{r\ell }^{Schr}(t)$ leads to the decay rate of the master
equation which plays the role of the time-dependent relaxation matrix in the
Redfield equation (\ref{6}).

By inserting Eq. (\ref{10}) into Eq. (\ref{9}) and performing the trace over
the environments degrees of freedom, we obtain the master equation in the
interaction picture 
\begin{align}
\frac{d\sigma (t)}{dt}& ={\sum\limits_{r,r^{\prime }}}\{\mathcal{F}%
_{rr^{\prime }}(t)\left[ \mathcal{O}_{r^{\prime }}\sigma (t),\mathcal{O}%
_{r}^{\dagger }\right]  \notag \\
& +\mathcal{G}_{rr^{\prime }}(t)[\mathcal{O}_{r^{\prime }}^{\dagger
}\sigma(t),\mathcal{O}_{r}]+H.c.\},  \label{11}
\end{align}
where we have defined the functions 
\begin{align*}
\mathcal{F}_{rr^{\prime }}(t)& =2\lim_{\tau \rightarrow 0}\left[ \frac{1}{%
\tau }\int_{t}^{t+\tau }dx\int_{t}^{x}dx^{\prime }\left\langle
\Gamma_{r}^{\dagger }(x)\Gamma _{r^{\prime }}(x^{\prime })\right\rangle %
\right], \\
\mathcal{G}_{rr^{\prime }}(t)& =2\lim_{\tau \rightarrow 0}\left[ \frac{1}{%
\tau }\int_{t}^{t+\tau }dx\int_{t}^{x}dx^{\prime }\left\langle
\Gamma_{r}(x)\Gamma _{r^{\prime }}^{\dagger }(x^{\prime })\right\rangle %
\right] .
\end{align*}

For the environments considered in this work it follows that: $%
\left\langle\Gamma _{r}^{\dagger }(t)\Gamma _{r^{\prime }}^{\dagger
}(t^{\prime})\right\rangle =\left\langle \Gamma _{r}(t)\Gamma _{r^{\prime
}}(t^{\prime})\right\rangle =0$, $\left\langle \Gamma _{r}^{\dagger
}(t)\Gamma_{r^{\prime }}(t^{\prime })\right\rangle =\left\langle
n_{r}\right\rangle\delta _{rr^{\prime }}$ and $\left\langle \Gamma
_{r}(t)\Gamma _{r^{\prime}}^{\dagger }(t^{\prime })\right\rangle =\left(
\left\langle n_{r}\right\rangle +1\right) \delta _{rr^{\prime }}$, $%
\left\langle n_{r}\right\rangle $ being the thermal average excitation of
the $r$th environment. These relations, of course, depend on the state of
the environment. We observe that this master equation describes the
Markovian evolution of a general time-dependent system, provided that the
conditions $i)$ and $ii)$ of the last section are satisfied.

\section{The characteristic relaxation times}

In this section, restricting us to the case of spin-$1/2$ systems, we aim to
derive the Bloch equations for the evolution of the magnetization components
of $N$ non-interacting spins. First, we obtain the Bloch equations from the
Redfield formalism, relating the characteristic relaxation times with the
properties of the associated classical stochastic environment. Next,
computing the evolution of the average magnetization from the master
equation formalism, we are able to link the characteristic relaxation times
with the properties of the quantum environment.

\subsection{From the Redfield to the Bloch equations}

Let us consider here a spin-$1/2$ system placed in a constant magnetic field
in $z$ direction. The frequency gap between the two Zeeman levels defines
the Larmor frequency $\omega _{L}\left( t\right) =\omega _{1}\left(
t\right)-\omega _{0}\left( t\right)$, with $\omega _{1}\left( t\right)$ and $%
\omega _{0}\left( t\right)$ representing the frequencies of the excited and
the ground state, respectively. The modulation of these frequencies are due
to some external influence, like an additional time-dependent magnetic
field. The bare Hamiltonian of the spin-$1/2$ system is then given by $%
H_{S}(t)=\omega _{L}\left( t\right) I_{z}$. The action of the environment
over the system is modelled by the spin-lattice Hamiltonian 
\begin{equation}
H_{SL}(t)=-\gamma _{n}\sum_{q}\lambda _{q}(t)I_{q},  \label{12}
\end{equation}
where $\gamma _{n}$ is the gyromagnetic factor, $q$ labels the orthogonal
Cartesian directions $\{x,y,z\}$, $\lambda _{q}(t)$ refers to the lattice
stochastic fluctuation in $q$ direction, and $I_{q}=\sigma _{q}/2$ stands
for the spin (Pauli) operator.

For this system, the spectral density given in Eqs. (\ref{8}) becomes 
\begin{equation}
J_{kn,n^{\prime }k^{\prime }}(t,\Omega _{n^{\prime
}k^{\prime}})=\sum\limits_{q}I_{q}^{kn}I_{q}^{n^{\prime }k^{\prime
}}\Theta_{q}(t,\Omega _{n^{\prime }k^{\prime }}),  \label{12a}
\end{equation}
where $I_{q}^{kn}=\left\langle k\right\vert I_{q}\left\vert n\right\rangle$
and 
\begin{equation}
\Theta _{q}(t,\Omega _{n^{\prime }k^{\prime }})=\gamma
_{n}^{2}\lambda_{q}^{2}\int_{0}^{t}dt^{\prime }e^{-|t^{\prime
}|/\tau_{0}}e^{i\Omega _{n^{\prime }k^{\prime }}(t+t^{\prime })}.
\label{12b}
\end{equation}
To derive Eq. (\ref{12b}) we have assumed isotropic stochastic fluctuations 
\cite{Slichter}, by which 
\begin{equation*}
\overline{\lambda _{q}(t)\lambda _{q^{\prime }}(t+t^{\prime })}%
=\delta_{qq^{\prime }}\lambda _{q}^{2}e^{-|t^{\prime }|/\tau _{0}},
\end{equation*}
$\lambda _{q}^{2}$ being a mean value depending on the specific nature of
the spin system and $\tau _{0}$ the environment correlation time, measuring
the rate of flips between the bath spins due to a specific anisotropic spin
interaction (chemical shift, dipolar coupling, etc. ) \cite{Anderson}. Note
that we have assumed that the mean values of the coupling $\lambda _{q}(t)$
are not affected due to the time-dependence of the system. This is quite
reasonable since we are modelling the environment as a stochastic noise
source. Remembering Sec. II, we have 
\begin{equation*}
\Omega _{kn}(t)=\int_{0}^{t}d\tau \left[ \omega _{k}\left( \tau \right)
-\omega _{n}\left( \tau \right) \right] ,
\end{equation*}
which is just the integral of the Larmor frequency with a positive or
negative signal, depending on the difference $k-n$ ($k,n=0,1$).

Next, by substituting Eq. (\ref{12a}) into Eq. (\ref{7}), we obtain the
elements of the relaxation matrix 
\begin{align}
R_{kn,n^{\prime }k^{\prime }}(t)=& \sum_{q}\left\{ \left[ \Theta_{q}(t,%
\Omega _{n^{\prime }k^{\prime }})e^{i\Omega _{kn}(t)}\right. \right.  \notag
\\
& \left. +\Theta _{q}(t,\Omega _{kn})e^{i\Omega _{n^{\prime }k^{\prime}}(t)} 
\right] I_{q}^{kn}I_{q}^{n^{\prime }k^{\prime }}  \notag \\
& -\sum_{j}\left[ \delta _{k^{\prime }n^{\prime}}I_{q}^{kj}I_{q}^{jn}\Theta
_{q}(t,\Omega _{jn})e^{i\Omega _{kj}(t)}\right.  \notag \\
& \left. \left. +\delta _{kn}I_{q}^{jk^{\prime }}I_{q}^{n^{\prime
}j}\Theta_{q}(t,\Omega _{n^{\prime }j})e^{i\Omega _{jk^{\prime }}(t)}\right]
\right\},  \label{13}
\end{align}
which enable us to compute the evolution of the mean value of the
magnetization $\left\langle I_{d}\right\rangle $ in an arbitrary $d$
direction: 
\begin{equation*}
\frac{d\left\langle I_{d}\right\rangle }{dt}=\frac{d}{dt}\mbox{Tr}\left[%
I_{d}\Sigma (t)\right] =\mbox{Tr}\left[ I_{d}\frac{d\Sigma (t)}{dt}\right].
\end{equation*}
By replacing Eqs. (\ref{6}) and (\ref{13}) into the r.h.s. of the last
equation, we obtain 
\begin{align*}
\frac{d\left\langle I_{d}\right\rangle }{dt}& =
-i\sum_{l,m}I_{d}^{lm}\left\langle m\right\vert \left[H_{S}(t),\Sigma \right]
\left\vert l\right\rangle \\
& +\sum_{q}\sum\limits_{l,m}\Theta _{q}(t,\Omega
_{ml})e^{-i\Omega_{ml}}I_{d}^{ml}\left\langle l\right\vert \left[ \left[
I_{d},I_{q}\right],\Sigma \right] \left\vert m\right\rangle .
\end{align*}
Since, for spin-$1/2$ systems, $\left\vert \Omega _{ml}(t)\right\vert
=\Omega (t)(1-\delta _{ml})$, the above equation can be separated into the
longitudinal and transverse field components, 
\begin{align*}
\frac{d\left\langle I_{d}\right\rangle }{dt}=&\sum\limits_{q=\{x,y\}}\sum%
\limits_{l,m}\kappa _{q}(t,\Omega_{ml})I_{d}^{ml}\left\langle l\right\vert %
\left[ \left[ I_{d},I_{q}\right],\Sigma \right] \left\vert m\right\rangle \\
& +\kappa _{z}\mbox{Tr}\left\{ I_{z}\left[ \left[ I_{d},I_{z}\right],\Sigma %
\right] \right\} ,
\end{align*}
where $\kappa _{q}(t,\Omega _{ml})=\Theta _{q}(t,\Omega
_{ml})e^{-i\Omega_{ml}(t)}$. We stress that we have neglected the
free-evolution term in the above equation because we are only interested in
the effect of the environment induced dynamics. We also note that $%
\left\langle m\right\vert I_{x(y)}\left\vert l\right\rangle \neq 0$ if and
only if $m\neq l$ and $\left\langle m\right\vert I_{z}\left\vert
l\right\rangle \neq 0$ when $m=l$, which explains why the term $\kappa
_{x(y)}(t,\Omega _{ml})$ is a time-dependent function while $\kappa _{z}$ is
a constant. Writing the last equation in terms of the longitudinal and
transversal magnetizations, defined as $M_{z}=\left\langle
I_{z}\right\rangle $ and $\mathbf{M}_{\perp}=\left\langle I_{x}\right\rangle 
\hat{x}+\left\langle I_{x}\right\rangle \hat{y}$, respectively, we obtain 
\begin{align*}
\frac{dM_{z}(t)}{dt}& =-\mbox{Re}\left[ \kappa _{x}(t,\Omega
)+\kappa_{y}(t,\Omega )\right] \left\{ M_{z}(t)-M_{0}\right\} , \\
\frac{d\mathbf{M}_{\perp }(t)}{dt}& =-\frac{1}{2}\mbox{Re}\left[
\kappa_{x}(t,\Omega )+\kappa _{y}(t,\Omega )+2\kappa _{z}\right] \mathbf{M}%
_{\perp}(t),
\end{align*}
$M_{0}=\left\langle I_{z}\right\rangle _{eq}=\mbox{Tr}\left\{
I_{z}\sigma_{eq}\right\} $ being the equilibrium longitudinal magnetization.
Now, comparing these results with the phenomenological Bloch equations 
\begin{subequations}
\label{14}
\begin{align}
\frac{dM_{z}(t)}{dt}& =\frac{1}{T_{1}}\left\{ M_{z}(t)-M_{0}\right\} ,
\label{14a} \\
\frac{d\mathbf{M}_{\perp }(t)}{dt}& =-\frac{\mathbf{M}_{\perp }(t)}{T_{2}},
\label{14b}
\end{align}%
the characteristic relaxation times, in terms of the time-dependent decay
rates $\kappa _{q}$ in the classical stochastic environment, is defined as 
\end{subequations}
\begin{subequations}
\label{15}
\begin{align}
\frac{1}{T_{1}}& \equiv \mbox{Re}\left[ \kappa _{x}(t,\Omega
)+\kappa_{y}(t,\Omega )\right],  \label{15a} \\
\frac{1}{T_{2}}& \equiv \frac{1}{2}\mbox{Re}\left[ \kappa
_{x}(t,\Omega)+\kappa _{y}(t,\Omega )\right] +\kappa _{z},  \label{15b}
\end{align}
Equations (\ref{14}) and (\ref{15}) show that, in contrast to the
longitudinal rate $T_{1}$, the transverse decay rate $T_{2}$ is related to
an energy conserving process, affecting only the quantum coherence of the
system. This fact justify the choice of the amplitude- and phase-damping
channels for the quantum description of the spin system. It is worth
mentioning that $T_{1}$ can be controlled through the time-dependent
parameter $\Omega (t)$ while $T_{2}$ can only be partially controlled since
the decay rate $\kappa _{z}$ does not depend on $\Omega (t)$. Finally, from
Eqs. (\ref{15}) we obtain the well known relation between both
characteristic times 
\end{subequations}
\begin{equation}
\frac{1}{T_{2}}=\frac{1}{2T_{1}}+\kappa _{z}.  \label{16}
\end{equation}

\subsection{From the master equation to the Bloch equations}

In this section we consider the same system as before, but instead of a
classical noise the spin system interacts with two quantum environments. In
order to compute the evolution of the average magnetization from the master
equation (\ref{11}), which takes into account the amplitude- and
phase-damping channels, we first address the decay rates $\mathcal{F}%
_{rr^{\prime }}(t)$ and $\mathcal{G}_{rr^{\prime }}(t)$ defined in the end
of Sec. III. When considering the amplitude-damping channel ($r=a$) we
associate $\mathcal{O}_{a}$ ($\mathcal{O}_{a}^{\dagger }$) with the lowering
(raising) spin operators whereas in the case of phase-damping ($r=p$) we
define $\mathcal{O}_{p}$ as the Hermitian number excitation operator. We
then set $\mathcal{O}_{a}=I_{-}$ and $\mathcal{O}_{p}=I_{z}$. For both cases
we set the mean values for the environment operators $\left\langle
a_{rl}^{\dagger }a_{rk}\right\rangle =\left\langle
n_{r,k}\right\rangle\delta _{lk}$ and, consequently, $\left\langle
a_{rk}a_{rl}^{\dagger}\right\rangle =\left( \left\langle
n_{r,k}\right\rangle +1\right) \delta_{lk}$, $\left\langle
n_{r,k}\right\rangle $ being the thermal average excitation of the $k$th
mode of the $r$th environment. Considering that the environment frequencies
are very closely spaced to allow a continuum summation, such that $%
\sum\nolimits_{\ell }\rightarrow \left(2\pi
\right)^{-1}\int\nolimits_{-\infty }^{\infty }d\nu J_{r}(\nu )$, $J\left(
\nu\right) $ being the spectral density of the environment, we obtain, for
the amplitude damping case, the effective time-dependent decay rates 
\begin{subequations}
\label{17}
\begin{align}
\mathcal{F}_{a}(t)& =\frac{\left\langle n_{a}\right\rangle }{2\pi }%
\Theta_{a}(t),  \label{17a} \\
\mathcal{G}_{a}(t)& =\frac{\left( \left\langle n_{a}\right\rangle +1\right) 
}{2\pi }\Theta _{a}(t),  \label{17b}
\end{align}
where, remembering the time-dependence of the system-environment coupling in
the Schr\"{o}dinger picture $\gamma _{r\ell }^{Schr}(t)$, 
\end{subequations}
\begin{align*}
\Theta _{a}(t)& =\lim_{\tau \rightarrow 0}\left[ \frac{1}{\tau }%
\int_{t}^{t+\tau }dx\int_{t}^{x}dx^{\prime }\int_{-\infty }^{\infty }d\nu
J_{a}\left( \nu \right) \right. \\
\times & \left. \gamma _{a}^{Schr}(\nu ,x^{\prime })\gamma_{a}^{Schr}(\nu,x)%
\mbox{e}^{i\left[\Omega(x^{\prime })-\Omega (x)+\nu \left(x-x^{\prime
}\right) \right] }\right] .
\end{align*}
To obtain the effective decay rates in Eq. (\ref{17}) it was assumed, as
usual, that the thermal average excitation of the environment modes vary
slowly around the range of variation of the spin system frequency. This is a
good approximation when the environment is in a thermal state, as the
present case \cite{SZ}.

For the phase-damping channel, the effective decay rates are given by 
\begin{subequations}
\label{18}
\begin{align}
\mathcal{F}_{p}(t)& =\frac{\left\langle n_{p}\right\rangle }{2\pi }%
\Theta_{p}(t),  \label{18a} \\
\mathcal{G}_{p}(t)& =\frac{\left( \left\langle n_{p}\right\rangle +1\right) 
}{2\pi }\Theta _{p}(t),  \label{18b}
\end{align}
where we defined 
\end{subequations}
\begin{align*}
\Theta _{p}(t)=& \lim_{\tau \rightarrow 0}\left[ \frac{1}{\tau }%
\int_{t}^{t+\tau }dx\int_{t}^{x}dx^{\prime }\int_{-\infty }^{\infty }d\nu
J_{p}\left( \nu \right) \right. \\
& \times \left. \gamma _{p}^{Schr}(\nu ,x^{\prime })\gamma _{p}^{Schr}(\nu,x)%
\mbox{e}^{i\nu (x^{\prime }-x)}\right] .
\end{align*}

Due to the diagonal system-environment coupling associated with the
phase-damping case, we are able to compute $\Theta _{p}(t)$ without having
the explicit form of $\Omega (t)$. In fact, assuming that the spectral
density $J_{p}\left( \nu \right)$ as well as the system-environment coupling 
$\gamma _{p}^{Schr}(\nu ,x^{\prime })$ vary slowly around $\nu =0$, we
obtain the time-independent parameter 
\begin{align*}
\Theta _{p}& =\lim_{\tau \rightarrow 0}\left[ \frac{J_{p}\left( 0\right) }{%
\tau }\int_{t}^{t+\tau }\left[ \gamma _{p}^{Schr}(0,x)\right] ^{2}dx\right]
\\
& =J_{p}\left[ \gamma _{p}^{Schr}\right]^{2}.
\end{align*}
From this result we see that, in contrast to the case of the
amplitude-damping channel, the decay rates $\mathcal{F}_{p}$ and $\mathcal{G}%
_{p}$ do not acquire a time dependence due to the modulation of the system
frequency, resembling the result obtained in the Redfield formalism.
Finally, the master equation (\ref{11}) becomes 
\begin{align}
\frac{d\sigma (t)}{dt}=& \frac{\left\langle n_{a}\right\rangle }{2\pi }%
\Theta _{a}(t)\left[ I_{-}\sigma (t),I_{+}\right]  \notag \\
& +\frac{\left( \left\langle n_{a}\right\rangle +1\right) }{2\pi }%
\Theta_{a}(t)\left[ I_{+}\sigma (t),I_{-}\right]  \notag \\
& +\frac{\Theta _{p}}{2\pi }\left( 2\left\langle n_{p}\right\rangle
+1\right) \left[ I_{z}\sigma (t),I_{z}\right] +H.c.  \label{EqMestra}
\end{align}

Computing the evolution of the mean value of the magnetization $\left\langle
I_{d}\right\rangle$ in an arbitrary $d$ direction we obtain for the
longitudinal and transversal magnetizations 
\begin{align*}
\frac{dM_{z}}{dt} & =-\frac{\mbox{Re}\Theta_{a}(t)}{\pi}\left( 2\left\langle
n_{a}\right\rangle +1\right) M_{z}, \\
\frac{d\mathbf{M}_{\perp}}{dt} & =-\left[ \frac{\mbox{Re}\Theta _{a}(t)}{2\pi%
}\left( 2\left\langle n_{a}\right\rangle +1\right) \right. \\
& \left. +\frac{\mbox{Re}\Theta_{p}}{\pi}\left( 2\left\langle
n_{p}\right\rangle +1\right) \right] \mathbf{M}_{\perp}.
\end{align*}

As in the preceding subsection, we compose these equations with the Bloch
equations (\ref{14}), to obtain 
\begin{subequations}
\label{T}
\begin{align}
\frac{1}{T_{1}} & \equiv \frac{\mbox{Re}\Theta _{a}(t)}{\pi }%
\left(2\left\langle n_{a}\right\rangle +1\right) ,  \label{Ta} \\
\frac{1}{T_{2}} & \equiv \frac{\mbox{Re}\Theta _{a}(t)}{2\pi }%
\left(2\left\langle n_{a}\right\rangle +1\right) +\frac{\mbox{Re}\Theta _{p}%
}{\pi } \left( 2\left\langle n_{p}\right\rangle +1\right) ,  \label{Tb}
\end{align}
and, consequently, the relation 
\end{subequations}
\begin{equation*}
\frac{1}{T_{2}}=\frac{1}{2T_{1}}+\frac{\mbox{Re}\Theta _{p}}{\pi }%
\left(2\left\langle n_{p}\right\rangle +1\right) ,
\end{equation*}
which has the same structure as Eq. (\ref{16}), obtained by the Redfield
formalism. From Eqs. (\ref{15}) and (\ref{T}) we can make the
identifications 
\begin{subequations}
\label{19}
\begin{align}
\frac{\mbox{Re}\Theta _{a}(t)}{\pi }\left( 2\left\langle n_{a}\right\rangle
+1\right) & \equiv \mbox{Re}\left[ \kappa _{x}(t,\Omega
)+\kappa_{y}(t,\Omega )\right] ,  \label{19a} \\
\frac{\mbox{Re}\Theta _{p}}{\pi }\left( 2\left\langle n_{p}\right\rangle
+1\right) & \equiv \kappa _{z}.  \label{19b}
\end{align}
These equations show the connections between the semi-classical and quantum
approaches to open system dynamics. In the next two sections we will
construct the Kraus and the phenomenological operators for the
time-dependent system studied in this section.

\section{The operator-sum representation}

It is well known that every transformation that is given by a completely
positive map admits a representation of the form \cite{Cirac} 
\end{subequations}
\begin{equation}
\sigma (t)=\sum_{k}E_{k}\left( t\right) \sigma (0)E_{k}^{\dagger
}\left(t\right),  \label{20}
\end{equation}
with the Kraus operators $E_{k}\left( t\right) $ satisfying the following
relation 
\begin{equation*}
\sum_{k}E_{k}^{\dagger }\left( t\right) E_{k}\left( t\right) = 1.
\end{equation*}
Our goal in this section is to construct the operators $E_{k}\left( t\right)$
for both channels studied in the last section. To achieve this we will
consider the density operator evolution equations, which follows from the
Redfield or the master equation formalisms. Then, we compare these equations
with those shown in Eq. (\ref{20}) to obtain the time dependence of the
Kraus operators.

In the next two subsections we will adopt the basis defined in Sec. IV, that
diagonalize the spin operator $I_{z}$, $\left\{ \left\vert 0\right\rangle
,\left\vert 1\right\rangle \right\} $. Since both channels studied here are
independent, let us then consider each one of them separately.

\subsection{Phase-damping channel}

Considering only the phase-damping ($\Theta _{a}=0$), the master equation (%
\ref{EqMestra}) leads us to the following set of differential equations
satisfied by the elements of the density operator 
\begin{subequations}
\label{21}
\begin{align}
\frac{d\sigma _{11}(t)}{dt}& =0,  \label{21a} \\
\frac{d\sigma _{00}(t)}{dt}& =0,  \label{21b} \\
\frac{d\sigma _{10}(t)}{dt}& =-\Gamma _{p}\sigma _{10}(t),  \label{21c} \\
\frac{d\sigma _{01}(t)}{dt}& =-\Gamma _{p}\sigma _{01}(t),  \label{21d}
\end{align}
where we have defined $\Gamma _{p}=2\mbox{Re}\left[ \mathcal{F}_{p}+\mathcal{%
G}_{p}\right] =\mbox{Re}\Theta _{p}\left( 2\left\langle n_{p}\right\rangle
+1\right) /\pi $. Note that, as expected, the populations are not affected
by this noisy channel. We assume that the Kraus operators for this case are
given by \cite{Nielsen} 
\end{subequations}
\begin{subequations}
\label{22}
\begin{align}
E_{0}^{p}& =\sqrt{1-p\left( t\right) } 
\begin{pmatrix}
1 & 0 \\ 
0 & 1%
\end{pmatrix}%
,  \label{22a} \\
E_{1}^{p}& =\sqrt{p\left( t\right) } 
\begin{pmatrix}
1 & 0 \\ 
0 & -1%
\end{pmatrix}%
,  \label{22b}
\end{align}
with $p\left(t\right)$ being the parameter to be determined. Starting from
the initial density operator 
\end{subequations}
\begin{subequations}
\begin{equation*}
\sigma (0)= 
\begin{pmatrix}
\sigma _{11}^{0} & \sigma _{10}^{0} \\ 
\sigma _{01}^{0} & \sigma _{00}^{0}%
\end{pmatrix}%
,
\end{equation*}
and substituting Eq. (\ref{22}) into Eq. (\ref{20}), we thus obtain 
\end{subequations}
\begin{equation}
\begin{pmatrix}
\sigma _{11}(t) & \sigma _{10}(t) \\ 
\sigma _{01}(t) & \sigma _{00}(t)%
\end{pmatrix}
= 
\begin{pmatrix}
\sigma _{11}^{0} & \left( 1-2p\right) \sigma _{10}^{0} \\ 
\left( 1-2p\right) \sigma _{01}^{0} & \sigma _{00}^{0}%
\end{pmatrix}%
.  \label{DO}
\end{equation}
By imposing that the time derivative of the elements of the above evolved
density operator must be identical to those in Eqs. (\ref{21}), we derive
the following differential equation for the parameter $p\left( t\right)$: 
\begin{equation*}
\frac{dp}{dt}=-\frac{\Gamma _{p}}{2}\left( 2p-1\right),
\end{equation*}
with the initial condition $p\left( 0\right) =0$, arising from the fact that
the Kraus operators must reduce to the identity at the initial time. The
solution for $p\left( t\right) $ is thus given by 
\begin{equation}
p\left( t\right) =\frac{1}{2}\left\{ 1-\exp \left[ -\Gamma _{p}t\right]%
\right\},  \label{23}
\end{equation}
which finally defines the Kraus operators in Eqs. (\ref{22}).

\subsection{Amplitude-damping channel}

For the case of amplitude damping we assume that the Kraus operators are
given by \cite{Nielsen} 
\begin{subequations}
\label{24}
\begin{align}
E_{0}^{a}& =\sqrt{\gamma _{T}} 
\begin{pmatrix}
1 & 0 \\ 
0 & \sqrt{1-a\left( t\right) }%
\end{pmatrix}%
,  \label{24a} \\
E_{1}^{a}& =\sqrt{\gamma _{T}} 
\begin{pmatrix}
0 & \sqrt{a\left( t\right) } \\ 
0 & 0%
\end{pmatrix}%
,  \label{24b} \\
E_{2}^{a}& =\sqrt{1-\gamma _{T}} 
\begin{pmatrix}
\sqrt{1-a\left( t\right) } & 0 \\ 
0 & 1%
\end{pmatrix}%
,  \label{24c} \\
E_{3}^{a}& =\sqrt{1-\gamma _{T}} 
\begin{pmatrix}
0 & 0 \\ 
\sqrt{a\left( t\right) } & 0%
\end{pmatrix}%
,  \label{24d}
\end{align}
where $\gamma _{T}=\exp \left[ -\beta E\right] /Z$ is the Boltzmann factor, $%
E$ being the energy gap of the spin-$1/2$ levels and $Z=1+\exp \left[ -\beta
E\right] $ the partition function. By analogy with the preceding subsection
our aim is to obtain the differential equation obeyed by the parameter $%
a\left( t\right) $. To this end, we compute the time evolution given by Eq. (%
\ref{20}), to obtain 
\end{subequations}
\begin{equation*}
\begin{pmatrix}
\sigma _{11}(t) & \sigma _{10}(t) \\ 
\sigma _{01}(t) & \sigma _{00}(t)%
\end{pmatrix}
= 
\begin{pmatrix}
\gamma _{T}a+q\left( 1-\sigma _{00}^{0}\right) & \sqrt{q}\sigma _{10}^{0} \\ 
\sqrt{q}\sigma _{01}^{0} & \left( 1-\gamma _{T}\right) a+q\sigma _{00}^{0}%
\end{pmatrix}%
,
\end{equation*}
with $q=1-a$. By imposing that the differential equations derived from the
above density operator must be identical to those obtained from the master
equation (\ref{EqMestra}), with $\Theta _{p}=0$, given by 
\begin{align*}
\frac{d\sigma _{11}(t)}{dt}& =2\mbox{Re}\left\{ \mathcal{F}%
_{a}(t)\sigma_{00}(t)-\mathcal{G}_{a}(t)\sigma _{11}(t)\right\} , \\
\frac{d\sigma _{10}(t)}{dt}& =-\left\{ \mathcal{F}_{a}^{\ast }(t)+\mathcal{G}%
_{a}(t)\right\} \sigma _{10}(t), \\
\frac{d\sigma _{01}(t)}{dt}& =-\left\{ \mathcal{F}_{a}(t)+\mathcal{G}%
_{a}^{\ast }(t)\right\} \sigma _{01}(t), \\
\frac{d\sigma _{00}(t)}{dt}& =-2\mbox{Re}\left\{ \mathcal{F}%
_{a}(t)\sigma_{00}(t)-\mathcal{G}_{a}(t)\sigma _{11}(t)\right\} ,
\end{align*}
we finally obtain 
\begin{align*}
\left[ \gamma _{T}-\sigma _{11}^{0}\right] \frac{da}{dt}=& -2\mbox{Re}%
\left\{ \mathcal{F}_{a}(t)+\mathcal{G}_{a}(t)\right\} \left(
\gamma_{T}a+q\sigma _{11}^{0}\right) \\
& +2\mbox{Re}\left\{ \mathcal{F}_{a}(t)\right\} , \\
\frac{da}{dt}\sigma _{10}^{0}=& 2\left\{ \mathcal{F}_{a}(t)+\mathcal{G}%
_{a}^{\ast }(t)\right\} q\sigma _{10}^{0}, \\
\frac{da}{dt}\sigma _{01}^{0}=& 2\left\{ \mathcal{F}_{a}^{\ast }(t)+\mathcal{%
G}_{a}(t)\right\} q\sigma _{01}^{0}.
\end{align*}
The fourth equation being identical to the first one. Adding the last two
equations and remembering that the initial conditions for the elements of $%
\sigma (0)$ are arbitrary, the following set of differential equations must
be satisfied 
\begin{align}
\frac{\sigma _{11}^{0}-\gamma _{T}}{2}\frac{da}{dt}=& \mbox{Re}\left\{%
\mathcal{F}_{a}(t)+\mathcal{G}_{a}(t)\right\} \left( \gamma
_{T}a+q\sigma_{11}^{0}\right)  \label{25a} \\
& +\mbox{Re}\left\{ \mathcal{F}_{a}(t)\right\} ,  \notag \\
\frac{da}{dt}=& 2\mbox{Re}\left\{ \mathcal{F}_{a}(t)+\mathcal{G}%
_{a}(t)\right\} q,  \label{25b}
\end{align}%
together with the condition $a(0)=0$. The solution of Eq. (\ref{25b}) is
readily obtained as 
\begin{equation}
a\left( t\right) =1-\exp \left[ -2\int_{0}^{t}\mbox{Re}\left\{ \mathcal{F}%
_{a}(\tau )+\mathcal{G}_{a}(\tau )\right\} d\tau \right],  \label{26}
\end{equation}
and its substitution into Eq. (\ref{25a}) leads to the relation 
\begin{equation*}
\left( \gamma _{T}-1\right) \mbox{Re}\left\{ \mathcal{F}_{a}(t)\right\}+%
\gamma _{T}\mbox{Re}\left\{ \mathcal{G}_{a}(t)\right\} =0,
\end{equation*}
which, together with Eqs. (\ref{17}), results in the identity 
\begin{equation*}
\left\langle n_{a}\right\rangle =\frac{\gamma _{T}}{1-2\gamma _{T}}=\frac{1}{%
e^{\beta E}-1}.
\end{equation*}
Equation (\ref{26}) satisfies both Eqs. (\ref{25a}) and (\ref{25b}), thus
being the desired solution.

It is now straightforward to see the connection between both \textit{ab-inito%
} approaches, the Redfield and the master equation, with the operator sum
representation. Back to the identifications in Eqs. (\ref{19}) and the
definition of the relaxation times in Eqs. (\ref{T}), we can rewrite the
time dependence of the Kraus operators for the phase-damping channel as 
\begin{equation*}
p\left( t\right) =\frac{1}{2}\left\{ 1-\exp \left[ -\frac{\kappa _{z}t}{2}%
\right] \right\} ,
\end{equation*}
and for the amplitude-damping channel as 
\begin{equation*}
a\left( t\right) =1-\exp \left[ -\int_{0}^{t}\frac{1}{T_{1}\left(
\tau\right) }d\tau \right] .
\end{equation*}

It is interesting to point out that within the operator sum formalism the
distinction between the decay of the coherences and the $T_{2}$ decay in NMR
systems becomes \textbf{evident. While} the density matrix coherences decay
is completely independent from the $T_{1}$ relaxation time, there is an
intrinsic dependence of $T_{2}$ with $T_{1}$ through Eq. (\ref{16}). Thus,
one can conclude that the phenomenological $T_{2}$ does not reflect only the
decay of the quantum coherences of the system.

Before applying this formalism to a specific situation, let us make some
remarks regarding a very closely related technique, the
phenomenological-operator approach \cite{Miled} also introduced to simplify
the treatment of dissipative quantum systems.

\section{The phenomenological-operator approach}

The phenomenological-operator approach is a technique equivalent to the
operator sum representation, but taking explicitly into account the state of
the environment together with those of the open quantum system. From the
phenomenological operators we automatically derive the Kraus operators and
vice-versa.

\subsection{Phase-damping channel}

First, let us consider the coupling of the spin-$1/2$ states to a
surrounding phase-damping environment, which can be described by the map 
\begin{subequations}
\label{OP}
\begin{align}
|0\rangle |\mathcal{E}\rangle & \rightarrow |0\rangle \hat{\mathcal{T}}%
_{00}^{(p)}|\mathcal{E}\rangle  \label{OPa} \\
|1\rangle |\mathcal{E}\rangle & \rightarrow |1\rangle \hat{\mathcal{T}}%
_{11}^{(p)}|\mathcal{E}\rangle +|1\rangle \hat{\mathcal{T}}_{10}^{(p)}|%
\mathcal{E}\rangle ,  \label{OPb}
\end{align}%
where $|\mathcal{E}\rangle $ denotes the initial state of the environment
and the operators $\hat{\mathcal{T}}$, acting on this state, account for the
system-environment coupling. Since a phase-damping channel does not exchange
energy with the system, we obviously have the identity operator $\hat{%
\mathcal{T}}_{00}^{(p)}=\mathbf{1}$. Regarding the excited initial state $%
|1\rangle $, it will remain as such, with or without an additional phase
shift relative to the ground state $|0\rangle $, due to the action of the
environment. In the later case, we must impose (after the computed master
equation (\ref{EqMestra}) with $\Theta _{a}=0$), the decaying probability $%
\hat{\mathcal{T}}_{11}^{(p)}=\mbox{e}^{-\Gamma _{p}t}\mathbf{1}$, with the
above defined rate $\Gamma _{p}=2\mbox{Re}\left[ \mathcal{F}_{p}+\mathcal{G}%
_{p}\right] $, remembering that the time-dependent frequency does not lead
to a time-dependent relation rate $\Gamma _{p}$. In the former case we have $%
\hat{\mathcal{T}}_{10}^{(p)}=\sum_{j}\mathbf{f}_{j}(t)\left( a_{j}^{\dagger
}+a_{j}\right) $, with $\mathbf{f}_{j}(t)$ giving the probability amplitude
for environment, described by the creation and annihilation operators $%
a_{j}^{\dagger }$ and $a_{j}$, respectively, to induce a phase shift on the
excited state of the system. Assuming that the state of the environment is
modified such that $\left\langle \mathcal{E}\right\vert \hat{\mathcal{T}}%
_{10}^{(p)}|\mathcal{E}\rangle =0$, we obtain, after normalization of the
state vector $|1\rangle |\mathcal{E}\rangle $, the relation $\sum_{j}|%
\mathbf{f}_{j}(t)|^{2}=1-\mbox{e}^{-2\Gamma _{p}t}$. It is straightforward
to verify that the map in Eq. (\ref{OP}) leads exactly to the density
operator (\ref{DO}) derived for the phase-damping process.

\subsection{Amplitude-damping channel}

Before addressing the non-zero temperature case, we first consider the
coupling of the spin-$1/2$ states to an amplitude-damping environment at $%
T=0 $ K, described by the map 
\end{subequations}
\begin{subequations}
\begin{align*}
|0\rangle |\mathcal{E}\rangle & \rightarrow |0\rangle \hat{\mathcal{T}}%
_{00}^{(a)}|\mathcal{E}\rangle , \\
|1\rangle |\mathcal{E}\rangle & \rightarrow |1\rangle \hat{\mathcal{T}}%
_{11}^{(a)}|\mathcal{E}\rangle +|0\rangle \hat{\mathcal{T}}_{10}^{(a)}|%
\mathcal{E}\rangle .
\end{align*}
With the environment in the vacuum state we obviously have the identity
operator $\hat{\mathcal{T}}_{00}^{(a)}=\mathbf{1}$, and after the computed
master equation (\ref{EqMestra}) with $\Theta _{p}=0$, we must impose that $%
\hat{\mathcal{T}}_{11}^{(a)}=\exp \left[ -2\int_{0}^{t}\mbox{Re}\left\{ 
\mathcal{F}_{a}(\tau )+\mathcal{G}_{a}(\tau )\right\} d\tau \right] \mathbf{1%
}$. For a time-independent system we obtain the usual solution $\hat{%
\mathcal{T}}_{11}^{(a)}=\mbox{e}^{-\Gamma _{a}t}\mathbf{1}$, where $\Gamma
_{a}=2\mbox{Re}\left[ \mathcal{F}_{a}+\mathcal{G}_{a}\right]$. For the
operator $\hat{\mathcal{T}}_{10}^{(a)}$ associated with the excitation of
one of the infinite environment modes, we have $\hat{\mathcal{T}}%
_{10}^{(a)}=\sum_{j}\mathbf{g}_{j}(t)a_{j}^{\dagger }$, with $\mathbf{g}%
_{j}(t)$ giving the probability amplitude for the excitation of the $j$th
oscillator mode of the environment. After normalization of the wave vector $%
|1\rangle |\mathcal{E}\rangle $, we obtain $\sum_{j}|\mathbf{g}%
_{j}(t)|^{2}=1-\mbox{e}^{-2\Gamma _{a}t}$.

By turning our attention to the case of a non-zero temperature environment,
we can write the extended map 
\end{subequations}
\begin{align*}
|0\rangle |\mathcal{E}\rangle & \rightarrow |0\rangle \hat{\mathcal{T}}%
_{00}^{(a)}|\mathcal{E}\rangle +|1\rangle \hat{\mathcal{T}}_{01}^{(a)}|%
\mathcal{E}\rangle , \\
|1\rangle |\mathcal{E}\rangle & \rightarrow |1\rangle \hat{\mathcal{T}}%
_{11}^{(a)}|\mathcal{E}\rangle +|0\rangle \hat{\mathcal{T}}_{10}^{(a)}|%
\mathcal{E}\rangle ,
\end{align*}
where, now, instead of the identity operator $\hat{\mathcal{T}}_{00}^{(a)}=%
\mathbf{1}$, we must account for the probability of excitation of the system
due to the environment background photons. Since the operator $\hat{\mathcal{%
T}}_{00}^{(a)}$ is associated with an event at which the environment is not
excited, it must remain proportional to the identity. Moreover, as far as
the probability for the system to remain in the ground state must decrease
in a rate proportional to the environment temperature, we naturally impose
that $\hat{\mathcal{T}}_{00}^{(a)}=\sqrt{1-(1-\mbox{e}^{-2\Gamma_{a}t})%
\gamma _{T}}\mathbf{1}$. From the above assumption for $\hat{\mathcal{T}}%
_{00}^{(a)}$ we straightforwardly obtain for $\hat{\mathcal{T}}%
_{01}^{(a)}=\sum_{j}\mathbf{h}_{j}(t)a_{j}$, the relation $\sum_{j}|\mathbf{h%
}_{j}(t)|^{2}=\sqrt{(1-\mbox{e}^{-2\Gamma _{a}t})\gamma _{T}}$, with $%
\mathbf{g}_{j}(t)$ giving the probability amplitude for the system to be
excited by the $j$th oscillator mode of the environment. Regarding the
operator $\hat{\mathcal{T}}_{11}$, we know that it must be also proportional
to the identity since the environment must remain unaffected. However, for
the case of non-zero temperature, the probability for the system to remain
in the excited state must decrease in a rate smaller than the exponential
decay factor $\mbox{e}^{-2\Gamma _{a}t/2}$ coming from an environment at
absolute zero. Moreover, the equilibrium probability must depend on the
thermal average photon number $\left\langle n_{a}\right\rangle $, such that
we impose $\hat{\mathcal{T}}_{11}^{(a)}=\sqrt{\mbox{e}^{-2\Gamma_{a}tt}+(1-%
\mbox{e}^{-2\Gamma _{a}tt})\gamma _{T}}\mathbf{1}$. Consequently, for the
complementary operator $\hat{\mathcal{T}}_{10}^{(a)}=\sum_{j}\mathbf{\tilde{h%
}}_{j}(t)a_{j}^{\dagger }$ we obtain $\sum_{j}|\mathbf{\tilde{h}}%
_{j}(t)|^{2}=\sqrt{(1-\mbox{e}^{-\gamma t})(1-\gamma _{T})}$, $\mathbf{h}%
_{j}(t)$ being the probability amplitude for the excitation of the $j$th
oscillator mode of the environment at non-zero temperature.

\section{Coherence control}

In Ref. \cite{Lucas} it was proposed a method to circumvent the decoherence
process of a non-stationary system under amplitude-damping channel. In that
work, the authors focused on the state protection of a cavity mode whose
modulation of the frequency $\omega (t)$ was engineered through the
atom-field interaction. The master equation approach was used to investigate
the dynamics of the cavity mode, assuming its interaction with the
environment to be proportional to 
\begin{equation}
\frac{\xi ^{2}}{\left[ \omega (t)-\nu \right] ^{2}+\xi ^{2}},  \label{27}
\end{equation}
$\nu $ being the continuous frequency of the environment and the parameter $%
\xi $ accounting for the Lorentzian sharpness of the coupling around the
frequency $\omega (t)$. Thus the Lorentzian coupling, which is justified in
the system-environment weak-coupling regime, \textquotedblleft
follows\textquotedblright\ the evolution of the frequency of the system, as
expected under the sudden coupling approximation.

The modulation of the frequency was engineered to be of the form 
\begin{equation}
\omega (t)=\omega _{0}+\chi \sin \zeta t,  \label{28}
\end{equation}
with $\omega _{0}$ being the static frequency of the cavity mode. The
condition $\zeta /\omega _{0}\ll 1$, easily achieved within typical
experimental conditions as in NMR and cavity quantum electrodynamics, define
the adiabatic modulation of the frequency. Out of this regime, when $\zeta$ $%
\gtrsim \omega _{0}$ we reach the regime of the Casimir-like effect \cite%
{Lucas-CE}, where the decoherence mechanism of the cavity mode are
completely distinct from the one discussed here.

Within the condition $\zeta /\omega _{0}\ll 1$, we show that the control of
the decoherence process is achieved by means of the two parameters 
\begin{align}
\eta & \equiv \frac{\Gamma _{0}}{\zeta },  \label{29a} \\
\varepsilon & \equiv \frac{\xi }{\chi }\sim \frac{\Gamma _{0}}{\chi },
\label{29b}
\end{align}
where $\Gamma _{0}$ is the natural decay rate of the cavity mode and we have
assumed, as it is expected, that $\xi \sim \Gamma _{0}$. It was demonstrated
in Ref. \cite{Lucas} that a significant attenuation of the decoherence
occurs when both of these parameters are smaller than unit. This is seen
from the derived time-dependent decay rate of the cavity mode, which gets
weaker as one or both parameters $\eta$ and $\varepsilon$ decreases. The
physical reason for this can be seen as follows: The characteristic time
interval for an appreciable action of the environment over the stationary
system is around $\Gamma _{0}^{-1}$. However, when the frequency of the
system changes continuously, its rate of variation (proportional to $\zeta$)
plays a crucial role in the effective coupling between the system and the
environment. Remembering that this coupling occurs around $\omega (t)$, in a
region defined by the Lorentzian sharpness $\xi $ [see Eq. (\ref{27})], a
rate of variation $\zeta$ significantly larger than $\Gamma _{0}$, such that 
$\eta \ll 1$, makes difficult an effective action of the environment over
the system since their interaction time is reduced proportionally to $\eta $%
. Otherwise, when $\zeta $ is smaller than $\Gamma _{0}$, an effective
action of the environment takes place, inducing the relaxation of the system
before a significant change of its frequency. By its turn, the role of the
amplitude of the oscillation $\chi$ is to trigger the action of the rate of
variation $\zeta $. In fact, when the amplitude $\chi$ is smaller than the
Lorentzian sharpness $\xi$, the non-stationary system does not leave the
region (in frequency space) of its effective coupling with the environment,
thus decaying as a stationary system, whatever the value of $\zeta$.
However, when $\chi$ is larger than $\xi$, the effective system-environment
coupling moves to different regions of the spectrum, thus triggering the
action of the rate of variation $\zeta$ as described above \cite{Lucas}.

Now, we apply the same idea to the case of a spin-$1/2$ system considering
the Redfield formalism to treat the decoherence process. The same
conclusions of Ref. \cite{Lucas} are obtained here, \emph{but without
imposing any specific form for the system-environment coupling}, as in Eq. (%
\ref{27}), since a Lorentzian time-dependent effective decay rate for the
spin-$1/2$ system automatically appears from the Redfield formalism applied
to NMR systems. As far as we do not have to define the function in Eq. (\ref%
{27}), we stress that the parameter $\varepsilon $, which takes place
explicitly in the effective time-dependent decay rate of the cavity mode in
Ref. \cite{Lucas}, does not appear in the present spin-$1/2$ case. Instead,
we must use the ratio $\chi /\zeta $, also defined in Ref. \cite{Lucas},
weighting the contributions of parameters $\eta $ and $\varepsilon $.

The spin-$1/2$ system, interacting with the amplitude-damping environment,
is described by the Hamiltonian $H_{S}(t)=\omega _{L}\left( t\right) I_{z}$,
with $\omega _{L}\left( t\right) $ being modulated as in Eq. (\ref{28}). We
note that such a frequency modulation beras no conexion with To see how the
protocol works, consider the solution of the differential equation governing
the evolution of the longitudinal magnetization [see Eq. (\ref{14a})], which
decays with the effective rate given by 
\begin{equation*}
D\left( t\right) =\int\limits_{0}^{t}d\tau \frac{1}{T_{1}\left( \tau \right) 
},
\end{equation*}%
$T_{1}\left( t\right) $ defined in Eq. (\ref{15a}). Therefore, to circumvent
decoherence we have to make $T_{1}\left( t\right) $ grater than its static
value $T_{1}^{0}$ (equivalent to $\Gamma _{0}^{-1}$ in the previous case of
a damped cavity mode). This fact can also be directly seen from the
operators in Eqs. (\ref{24}), which reduces to the identity when $%
a(t)\rightarrow 0$, i.e., when $D\left( t\right) \rightarrow 0$. By
considering an isotropic and homogeneous environment, Eqs. (\ref{12b}), (\ref%
{15a}), and (\ref{28}) leads to the following expression for $D$ 
\begin{align}
D\left( \tau \right) =& 2\left( \gamma _{n}\lambda T_{1}^{0}\right)
^{2}\int\limits_{0}^{\tau }d\tau _{2}\int\limits_{0}^{\tau _{2}}d\tau
_{1}\exp \left\{ -\frac{T_{1}^{0}\tau _{1}}{\tau _{0}}\right\}   \notag \\
& \times \cos \left\{ T_{1}^{0}\omega _{L}\tau _{1}+2\frac{\chi }{\zeta }%
\sin ^{2}\left( 2\frac{\tau _{1}}{\eta }\right) \right\} ,  \label{30}
\end{align}%
where we have performed the change of variable $\tau =t/T_{1}^{0}$ and $\eta
=\left( T_{1}^{0}\zeta \right) ^{-1}$ as in Eq. (\ref{29a}). The
gyromagnetic factor $\gamma _{n}$ and the lattice stochastic fluctuation in $%
q$ direction $\lambda \equiv \lambda _{q}$ is defined in Eq. (\ref{12}).

In Fig. 1 we plot the function $a(\tau )=1-\mbox{e}^{-D(\tau )}$ against the
dimensionless time $\tau$, for some values of $\eta$ and the ratio $\chi
/\zeta $. As can be seen, the relaxation decay rate for the stationary case $%
\chi =0$, the black solid curve, is significantly attenuated when the
control parameter $\eta $ decreases and/or the ratio $\chi /\zeta $
increases. The red-dashed and the blue-dotted curves correspond to the pairs
of values ($\eta =0.1,\chi /\zeta =1$) and ($\eta=0.1,\chi /\zeta =10$),
respectively. The curves corresponding to the pairs ($\eta =0.01,\chi /\zeta
=1$) and ($\eta =0.01,\chi /\zeta =10$) are very close to the red dashed and
blue dotted lines, respectively, and, therefore, they are not shown in the
figure. From this fact we conclude that the role of parameter $\chi$ (the
amplitude of the modulation) in the decoherence control is more effective
than the role played by $\zeta$ (the modulation frequency). This result is
in perfect agreement with the one obtained in Ref. \cite{Lucas}. An
important observation is that this protocol is completely independent of the
temperature of the environment, as well as of the initial state of the
system.

\begin{figure}[htb]
\centering
\includegraphics[scale=0.41]{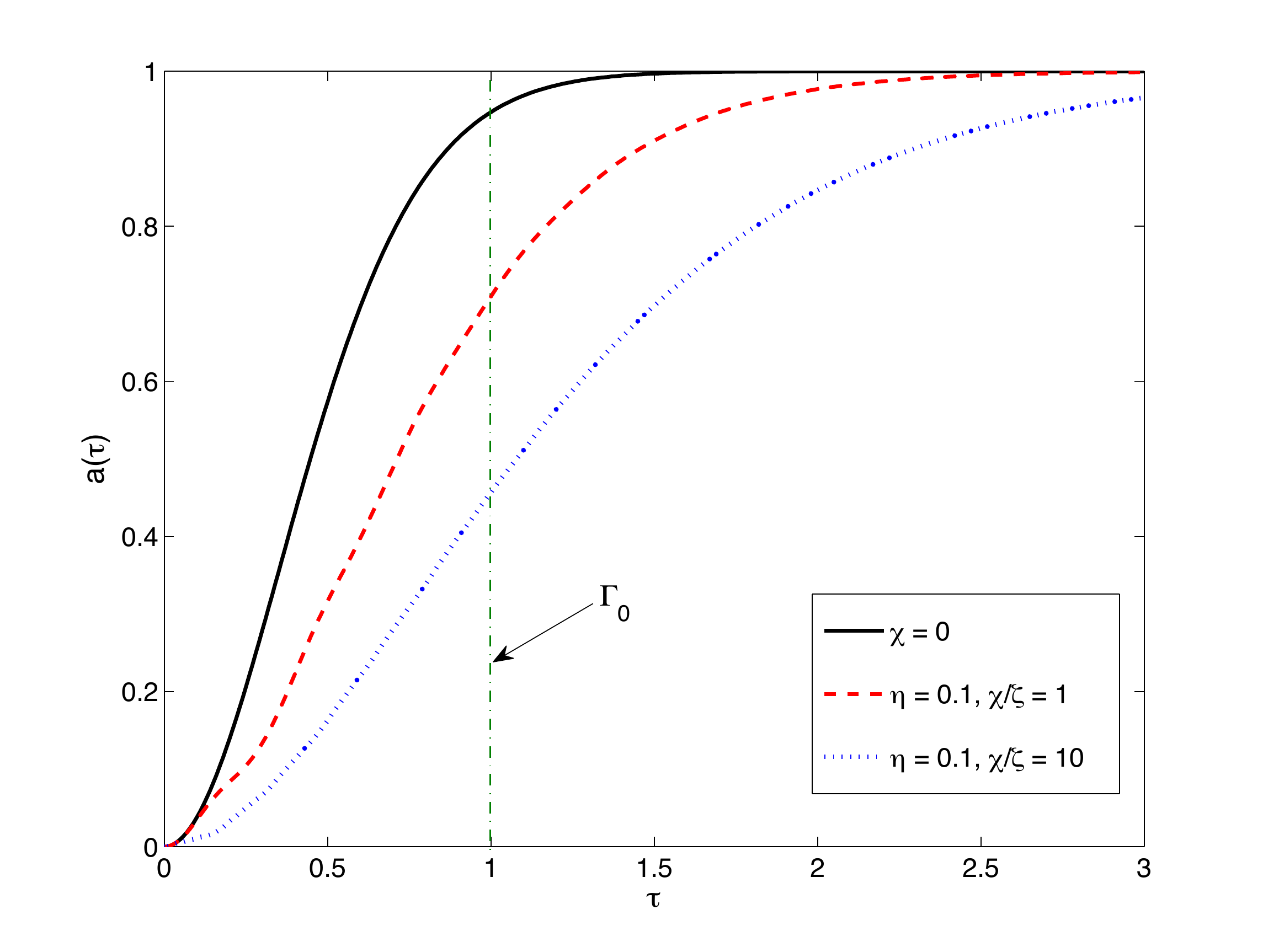}
\caption{(Color online) Plot of the decay function $a(t)$ defined in Eq. (%
\protect\ref{26}) for some values of the control parameters $\protect\eta $
and $\protect\chi /\protect\zeta $. The black solid line represents the
uncontrolled case, where $\protect\chi =0$. The other two shows the
effective control of the relaxation process due to the modulation of the
system frequency. The vertical green dash-dotted line marks the relaxation
time for the static (uncontrolled) case.}
\label{fig1}
\end{figure}

A possible implementation of the proposed scheme could be realized in an NMR
experiment as discussed below. In addition to the static Zeeman field $%
\mathbf{B}_{0}$ which defined the Larmor frequency $\omega
_{L}=\gamma_{n}B_{0}$, another parallel time-dependent field $\mathbf{B}(t)$
is applied to modulate the frequency of the spin-$1/2$ system. This
additional field may be furnished by a Helmholtz like coil surrounding the
probe, traversed by a tailored current which provides the time-dependent
component $\gamma _{n}B(t)$ of the frequency. By imposing the
time-dependence of the auxiliary field to be of the form%
\begin{equation*}
\gamma _{n}B(t)=\chi \sin \left( \zeta t\right),
\end{equation*}
we obtain the frequency modulation given by Eq. (\ref{28}). To circumvent
decoherence we must have $\eta$ $\ll 1$ [see Eq. (\ref{29a}) and Fig. 1],
which implies that the modulation frequency must obeys 
\begin{equation}
\zeta \gg \left( T_{1}^{0}\right) ^{-1}.  \label{33}
\end{equation}
This relation is in agreement with the adiabatic condition, since we have,
in general, $T_{1}\gg \omega _{L}^{-1}$. The other condition for the
protocol to work is that the amplitude $\chi$ of the oscillation of the
system frequency must be greater than the spectral sharpness of the coupling
which, as said before, is expect to be of the order of $\left(T_{1}^{0}%
\right) ^{-1}$, such that 
\begin{equation}
\chi \gg \left( T_{1}^{0}\right) ^{-1}.  \label{34}
\end{equation}

Besides the conditions (\ref{33}) and (\ref{34}), the magnetic field $B(t)$
must also obeys the adiabatic condition [see equation bellow Eq. (\ref{28}%
)], which leads to $\left\vert B(t)\right\vert \ll \left\vert
B_{0}\right\vert $. This conditions can be easily attained in a typical NMR
experiment. Considering a Hydrogen nuclei at $B_{0}\sim 10$ T ($\omega
_{L}\simeq 400$ MHz) and $T_{1}\sim 1$ s, we obtain, together with condition
(\ref{33}), the following range for the magnitude of the control field $B(t)$
for the protocol to work, i.e., $\left( \gamma _{n}T_{1}^{0}\right) ^{-1}\ll
\left\vert B(t)\right\vert \ll \left\vert B_{0}\right\vert $: 
\begin{equation*}
10^{-6}\text{ T}\ll \left\vert B(t)\right\vert \ll 10\text{ T.}
\end{equation*}

Note that the NMR setup is appropriate for this kind of experiment due to
the fact that the relaxation time is relatively large compared with other
platforms. This fact permit us to realize the experiment in a
high-temperature environment. It is also important to observe that despite
in the NMR experiments both amplitude- and phase-damping are present, there
exist experiments where only the amplitude-damping are probed, so they can
be used to check out our proposals. Therefore, we can only control the
relaxation time $T_{1}$, but not $T_{2}$.

\section{Final discussions}

We have presented here a unified view of the semiclassical Redfield
formalism and the quantum master equation approach for a time-dependent spin
system. Focusing on a spin-$1/2$ system, we shown the equivalence between
both approaches through the fact that they lead to the same Bloch equations
and, consequently, to the same characteristic longitudinal $T_{1}$ and
transversal $T_{2}$ relaxations times. We verified that only $T_{1}$ is
affected by the time-dependency of the system frequency and built the Kraus
and the phenomenological operators for the spin-$1/2$ system under both the
amplitude- and the phase-damping channels assumed within the master equation
approach.

As an application, we revisited a protocol to circumvent relaxation and,
consequently, the coherence control of a non-stationary system \cite{Lucas}.
In contrast to the protocol in Ref. \cite{Lucas}, in the present case, we do
not imposed a functional form for the system-environment coupling, which
emerges naturally from the Redfield semiclassical formalism. The coherence
control of the spin-$1/2$ system was demonstrated by enlarging the
longitudinal relaxation time through the modulation of the system frequency.
We stress that the protocol in Ref. \cite{Lucas}, translated here for the
spin-$1/2$ system is different from the dynamical decoupling methods
presented in literature \cite{DD}, where one must interfere in the system in
time scales less than the bath correlation time. The frequency modulation
technique takes advantage of the pre existing natural frequency of the
system, adding to it a small amplitude to achieve such time scales. We also
discussed the implementation of this protocol to control the longitudinal
relaxation time in the NMR context.

It is worth stressing that the development present here applies for
Markovian environments and adiabatic modulation of the required
time-dependent frequency. Therefore, an extension of the present development
for non-Markovian environments as well as for interacting time-dependent
systems would be desirable, since non-Markovian environments is present in
many promising platforms for quantum information processing such as photonic
crystals \cite{PC}. Moreover, time-dependent interacting system would
considerably enlarge the perspective of the present work for quantum
information purposes.

Although the entanglement is not usually present in the NMR system \cite%
{NMREntang}, there exists classical and quantum correlations that could be
useful for some quantum information protocols \cite{Correlations}. The
decoherence control exposed here can then be directly applied to protect
these correlations from the action of the amplitude-damping channel.
Finally, we stress that the theory presented here is completely general and
can be applied to different platforms for quantum information processing
such as cavity quantum electrodynamics, trapped ions, quantum dots, etc.
However, for this statement to be valid we must assume that all the
rotations needed to implement quantum information processing occur at the
beginning or at the end of the experiment; in between them, the system
evolves only under the action of the environments.

\textbf{Acknowledgments}

The authors acknowledge financial support from UFABC, CAPES and FAPESP. JM,
RMS and LCC also thank the financial support from UFABC. This work was
performed as part of the Brazilian National Institute of Science and
Technology for Quantum Information (INCT-IQ).

\end{document}